# An Algorithm for File Transfer Scheduling in Grid Environments

Alexandra Carpen-Amarie[1,2], Mugurel Andreica[1], Valentin Cristea[1]
[1]University "Politehnica" of Bucharest, [2]INRIA/IRISA, Rennes, France
Alexandra.Carpen-amarie@irisa.fr, {mugurel.andreica,valentin}@cs.pub.ro

## Abstract

*This paper addresses the data transfer scheduling problem for Grid environments, presenting a centralized scheduler developed with dynamic and adaptive features. The algorithm offers a reservation system for user transfer requests that allocates them transfer times and bandwidth, according to the network topology and the constraints the user specified for the requests. This paper presents the projects related to the data transfer field, the design of the framework for which the scheduler was built, the main features of the scheduler, the steps for transfer requests rescheduling and two tests that illustrate the system's behavior for different types of transfer requests.*

## 1. Introduction

Grid-based computing environments have become a critical requirement in large-scale scientific and engineering research and the access to distributed data has become as important as access to distributed computational resources. Data-intensive applications require that massive datasets are transferred between the machines involved in the computation. Examples of such applications include experimental analysis and simulations in scientific disciplines such as high-energy physics, climate modeling, earthquake engineering.

The need for a file transfer scheduler emerges from the data placement requirements of the data-intensive applications, which deal with the transfer of large datasets, necessary for the input or obtained as an output of computation jobs. These transfers can have different time constraints and priorities associated, that show the scheduler the moment when they are needed in order to optimize the cost of the running application.

The challenges raised by the need to guarantee a certain quality of service in a Grid environment can be summarized as follows[4]:

- Resource heterogeneity – data transfers are mainly influenced by network heterogeneity, networks used to interconnect resources being significantly different in terms of bandwidth and latency.
- Resource Non-dedication – the available bandwidth of network links can vary over time, especially when the different sites are interconnected using simple Internet links, that can be also employed for traffic unrelated to the system.
- Dynamic behavior - in a Grid environment, resources can dynamically join or leave the system, and these changes have to be automatically detected.
- Dynamic scheduling – the system cannot use a static scheduling algorithm, as the information regarding all the tasks that have to be processed cannot be available before the start of the first task. Therefore, a dynamic algorithm is required, as a result of the need of a permanently updated estimation of the system state and of the necessity to make instant decisions when a new task arrives.

There are only a few scheduling approaches proposed for the file transfer problem, and most of them use a very simple scheduling algorithm, a First Come First Served policy, when dealing with new transfer requests. The scheduler presented in this paper offers a wider range of options for the transfer tasks, from the different types of requests to the time constraints that can be associated with them.

## 2. Related work

This section presents an overview of the projects that address the data scheduling problem and iterates through the main aspects they cover.

Stork[9] is a specialized scheduler for data placement activities in Grid, which previously have been done either manually or by using simple scripts. They can be queued, scheduled, monitored and managed in a fault tolerant manner. Stork relies on a framework in which computational and data placement jobs are treated and scheduled differently by their corresponding schedulers, and the management and synchronization of both types of jobs is performed by higher level planners[9].

Stork deals with data placement jobs originating from the necessity to move data sets into and out of data processing applications. This model implies the existence of dependencies between data transfer tasks. These are managed by DAGMan (Directed Acyclic Graph Manager), a meta-scheduler that submits the data transfer jobs to Stork, scheduling them according to an user defined input file where one can specify all input data transfers, output data transfers, data processing, and dependencies. In conclusion, Stork does not address dynamic scheduling (data transfer jobs are known at the beginning of the scheduling process, being specified in the input file needed by DAGMan).

The open source Globus® Toolkit[11][12] is a fundamental enabling technology for the Grid, letting people share computing power, databases, and other tools securely online across corporate, institutional, and geographic boundaries without sacrificing local autonomy. The Globus Toolkit provides a number of components for doing data management: GridFTP[13] for high-performance and reliable data transport; RFT (Reliable File Transfer)[5] - for managing multiple transfers. The Reliable Transfer Service (RFT) is a web service that provides interfaces for controlling and monitoring third party file transfers using GridFTP servers.

The File Transfer Service (FTS)[7] is the lowest-level data movement service defined in the gLite architecture. It is responsible for moving sets of files from one site to another, allowing participating sites to control the network resource usage. It is designed for point to point movement of physical files.

Globus RFT and gLite[8] FTS both offer reliable and robust management of data movement in a grid environment, and represent a significant advance over direct client management of data transfers[10]. FTS and RFT use a First Come First Served policy when processing transfer tasks, according to submission times. The user does not have the possibility to specify more options regarding the scheduling process or to state different constraints associated with the transfer requests. The scheduler proposed in this paper addresses these issues, enabling the user to specify different parameters for his transfers, like the needed bandwidth and the desired start moment, but also giving him the opportunity to leave these options unspecified and let the scheduler decide their optimal values.

There are other data transfer schedulers that address the following data transfer scheduling issues: scheduling for transfers with time constraints and bandwidth reservations[1][6], and dynamic bandwidth resizing for running requests, based on their priorities[1]. These schedulers do not tackle rescheduling issues when a transfer request cannot be fulfilled or when a transfer takes longer than its allocated time period, and they force the user to state the bandwidth that his transfer will use, without the system's possibility to choose a suitable bandwidth. The system presented in this paper brings an improvement from this point of view, offering a rescheduling algorithm that allows a rejected transfer to modify other, less important transfers in order to fit into the desired time period.

The dynamic bandwidth resizing method[1] involves the bandwidth resizing for all the running requests, in spite of their priorities (every running request gets a slice of the available bandwidth directly proportional with the request's priority). This feature is modified in this paper in the following way: a request that does not fit into its user specified position can only modify or reschedule requests that have a lower priority. This approach prevents the system from resizing all the requests when a low priority transfer cannot be scheduled.

## 3. The Scheduling Framework

The framework is conceived as a set of distributed services, which are loosely coupled and have well defined roles[1]. The whole system revolves around MonALISA[2][3], and is using different components of the framework in order to collect the monitoring data, to store it, to access and visualize it (Figure 1).

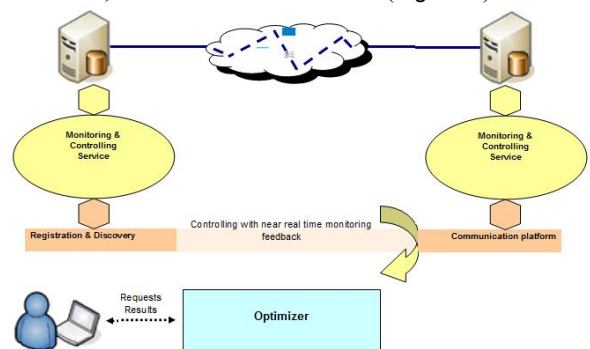

**Figure 1. System architecture**

The monitoring and controlling services are distributed throughout the system, on each end point where requests are actually executed, or close to intermediate points in the network that can provide monitoring information or that can be controlled. The interactions between the optimizer and the distributed services are assured by an advanced communication platform that, besides transporting monitoring and controlling messages, provides registration and discovery functions.

In general, a monitoring and controlling service is responsible for discovering the local network topology and retrieve monitoring information about the local nodes and links. It makes this information available to the optimizer through the flow of monitoring data. The optimizer, by retrieving this information from all the existing services, can infer the entire topology. The underlying details of each administrative domain are hidden from the optimizer, making the whole framework flexible and extensible. The optimizer service coordinates the entire system based on the monitoring information received from the other services and on the users' requests. The optimizer service is a high level client for the services below. It connects to the communication platform and discovers the available services. Then, using the monitoring and controlling interface, it receives the information about topology, requests, end hosts and issues the necessary commands for pursuing users' requests.

## 3.1. Network topology

The real network topology is abstracted as a graph within the optimizer, based on the available network segments reported by all the distributed services. This graph includes a set of nodes, interconnected through a number of links with additional information. The algorithms implemented in the optimizer run over this abstracted topology.

Among the parameters specific for each link the following are the most important, from the scheduler's point of view: *name*, *source and destination nodes*, *bandwidth*, *ID*.

## 3.2. Service requests

There are two fundamental scenarios in which a user interacts with the system: bandwidth reservation and data transfer. When a user wants to submit a transfer request, he has to select the type of request and its specific parameters. The system discovers one or more paths between the specified source and destination nodes and passes the paths on to the scheduler.

The scheduler must assign a start time, a duration and an allocated bandwidth (or just the ones that are not specified by the user) for every request, taking into account all the other requests previously added to the system (already running transfers and scheduled requests). Then, it returns to the user one or more offers obtained for the request in the scheduling process and the user must select the one that meets his needs.

For every type of request, a set of common attributes can be identified: *source and destination nodes*, *status*, *user and priority*, *time constraints*, *path* (the list of links chosen for the execution of this request), *bandwidth* (the maximal bandwidth used by the request. It can be either given by the user, or computed by the system, based on the individual bandwidth of each link in the path), *duration*, *monitoring parameters* (used bandwidth and finish status).

The possible statuses for a transfer request are: *offered* – an allocated, but not yet confirmed system response to user's request; *scheduled* – the system's response was accepted by the user and the request is scheduled for execution; *running* – the request is currently running; *finished* – the request has finished successfully; *error* – there was an error while setting up the path, or when running the request.

The user can specify for every request a desired time constraint:
- NONE: the request has no time constraints.
- ASAP: the request has to be scheduled as soon as possible - the scheduler finds the first time interval where the bandwidth needed by the request is available.
- NOT AFTER: the request cannot be scheduled after a specified time. The scheduler tries to add the request at the specified moment, provided that its bandwidth need is acceptable in the list of requests. If the request cannot fit at the desired time moment, it is rescheduled ASAP - the scheduled offer is returned to the client, if the begin time of the request is not after the user specified time.
- NOT BEFORE: the request cannot be scheduled before a user specified time moment. The system tries to schedule the request using as the transfer begin time the time specified by the user. If the request does not fit into the queue, it is rescheduled using a best fit policy at a moment after the one stated by the user.

## 4. The Transfer Scheduler

The scheduler keeps track of all the requests in the system. There is only one instance of the scheduler that is called when a user wants to make a transfer and it records all the paths used by the system, their allocated reservations, available bandwidths and available bandwidths per link.

The scheduler records all the paths that have associated transfer requests or running transfers. For every path, the system has to keep a list of the transfers scheduled on that path. Moreover, for every path, the system has to remember the available bandwidth, in order to know how much bandwidth it can allocate to a new request.

Initially, the bandwidth is computed as the minimum bandwidth of all the links contained in the path. When a new request is scheduled, the available bandwidth decreases during the time period allocated to the request. The outcome is that the scheduler has to record the different maximum available bandwidths over time.

The scheduler creates a list of time intervals where the bandwidth is constant. Every element of the list records the *beginning and the end time*, the *bandwidth* and a *list of requests* linked to this time interval.

When there are no requests, the maximum bandwidth of a path is computed as the minimum bandwidth of all the contained links. However, if one of these links also belongs to another path where there are scheduled requests, the link's bandwidth changes in time. As a result, every link needs to have a list of time intervals where the bandwidth is constant just like paths do. When a new request is added to the path, the available bandwidth of every link in the path changes as well. The links keep only the list of time intervals and associated bandwidths, and not the list of requests for every element. The request list is not necessary, as the list associated to a link is not used to schedule new requests, but only to compute the maximum available bandwidth for different paths that use the link.

## 5. Adding a new request

For every request type, the user can input different requirements for the scheduler: the desired bandwidth, the duration of the reservation, several types of time constraints. Each type of transfer goes through different steps on its way to being scheduled and then executed.

### 5.1. File Transfers

When a request for a simple transfer is submitted, the user has to provide the following information: the source node, the destination node, the source file and the file size. The scheduler handles in a different way the transfers with a specified bandwidth requirement and the transfers with an unspecified bandwidth.

The first case is when **the user specifies a bandwidth for the transfer**. The new transfer receives a prediction for the duration of the transfer, based on its size and the specified bandwidth. After that, the time constraint selected is checked. There can be four different configurations:

a) The user selected ASAP

The system selects from the list of time intervals the first subset of consecutive intervals where the request can fit given its bandwidth needs and duration. Next, the scheduler assigns a begin time to the request – the same as the begin time of the first interval in the selected list.

b) The user selected the NOT AFTER constraint and specified a begin time for the transfer

The transfer begin time is set to the moment specified by the user. Now the request has a begin time, a duration and a specified bandwidth. The system only finds the time interval that contains the begin time of the request and checks if enough bandwidth is available for the request for every interval in the list covered by the duration of the request, starting with the selected one.

If the required amount of bandwidth is available, the request is scheduled (added to the path's list) using these settings. Otherwise, the system tries to find a scheduling time for the request using an ASAP approach, provided that the begin time set by this method does not break the user specified time constraint (the begin time set is not after the time specified by the user).

c) The user selected the NOT BEFORE constraint and specified a begin time for the transfer

The transfer begin time is set to the moment specified by the user. The system attempts to find an appropriate place for the request in the same way it does for the NOT AFTER constraint. If this approach fails, the request is scheduled using the same method as for a request with no time constraints, checking that the selected start time complies with the constraint specified by the user (the start time is after the user specified time).

d) The user selected NONE – no time constraint

The system selects from the list of time intervals the last subset of consecutive intervals where the request can fit given its bandwidth needs and duration (excepting the last interval that always has the maximum bandwidth). Next, the scheduler assigns a begin time to the request – the same as the begin time of the first interval in the selected list.

The second type of file transfer is when **the user does not specify a bandwidth amount for his transfer**. The scheduler assigns the maximum possible bandwidth to the request. The request is scheduled like a request having a bandwidth specified by the user, according to the same four time constraint configurations. The estimated begin and end time are saved and the scheduler assigns a new bandwidth value for the request, equal to half of the initial value. The scheduling process is restarted and the two estimated finish times are compared. If the first result's end time is after the one from the second result, the latter is kept and scheduling process is once again repeated with a new, smaller value for the bandwidth. Otherwise, the first result is assigned to the request and offered to the user.

## 5.2. Bandwidth reservation transfers

For these transfers no actual data is transferred, but the user only wants to make sure that the selected link will have the required available bandwidth for the required duration. During that period, the user can use the link in any way as long as he does not exceed the specified amount of bandwidth. In contrast with the other transfer types, the user does not specify a certain file and file size. He just has to specify the source and destination nodes, the start and finish times and the amount of bandwidth he would like to have available in that period.

The scheduler's behavior is exactly the same as when a transfer request is placed using a NOT AFTER time constraint, because all the parameters are set by the user: bandwidth, start time and duration. The system tries to add the request at the specified time moment and, if the operation is not successful, the request is rescheduled using a ASAP approach and the result is offered to the user as an alternative to his specified constraints.

## 6. Request Rescheduling

There are two cases when a rescheduling of the existent requests is necessary: when a running transfer exceeds its allocated time period and overlaps other scheduled transfers or when a request having a high priority or specified time constraints cannot be scheduled on the selected path. The rescheduling process consists of three steps, performed first for the requests belonging to the current user and then for the requests belonging to other users:
1. The bandwidth modification for running requests.
2. The bandwidth modification for scheduled (not yet running) requests.
3. The rescheduling of the requests that can be moved and are not yet running.

## 6.1. Bandwidth modification for running requests

The requests that are suitable for this step of the rescheduling process must comply with the following rules:

- they have to be in one of the next two statuses: RUNNING, STARTING
- they must have a priority lower than the priority of the new request
- they must not have a bandwidth value specified by the user. The bandwidth has to be allocated by the scheduler.

The system tries to lower the used bandwidth for every request, until the new request can be scheduled into the desired position. A list consisting of requests that follow the above conditions is computed, having the elements sorted by several parameters: the priority (ascending sort), the size of the bandwidth modified for the request by subsequent rescheduling iterations (ascending sort), the size of the allocated bandwidth (ascending sort). For every running request, the system decreases the current bandwidth to the value that is the minimum between the needed bandwidth for the new request and half of the current bandwidth used by the request. The system tries to schedule again the new request. If the scheduling succeeds, the remaining requests will not be modified and will continue their execution with the same settings.

## 6.2. Bandwidth modification for scheduled requests

The requests that can be modified in this step of the rescheduling process must comply with the following rules:
- they have to be in the SCHEDULED status
- they must have a priority lower than the priority of the new request
- they must not have a bandwidth value specified by the user. The bandwidth has to be allocated by the scheduler.

For the scheduled requests, the approach is different, as when the bandwidth is modified, the duration of the request might change and affect other scheduled requests. As a consequence, all the requests that are suitable for this kind of modifications are first removed from the scheduler (not from the system and the transfer queues, because they will be reinserted in the scheduler into exactly the same position – they will have the same scheduled time – but with a different allocated bandwidth and possibly a different duration).

The system tries to schedule the new request. If the scheduling succeeds then the list of requests is reinserted into the scheduler, after they get modified. If the new request cannot be scheduled, the removed requests are reinserted into the scheduler after the new request goes through all the steps of the rescheduling process. If the request cannot be scheduled, they are reinserted unmodified, otherwise the system tries to add them to the scheduler using a Greedy algorithm approach.

The system sorts the modifiable requests, using the next rules: the priority (ascending sort), the size of the bandwidth modified for the request by subsequent rescheduling iterations (ascending sort),the size of the allocated bandwidth (descending sort). This sorting method is used to modify first the bandwidth for requests that have low priority, that were not modified a lot of times and that have a large amount of allocated bandwidth (these requests are more likely to free enough bandwidth for the new request and to allow the other requests to be scheduled unmodified). The list of requests is traversed from both directions. The system extracts one request from the beginning of the list, changes its bandwidth and then it schedules the modified request in the initial position. Then an element from the end of the list is processed. The system tries to add it to the scheduler unmodified. If the operation fails, the request will be scheduled like the others, with a modified bandwidth.

These two steps repeat until all the modifiable requests have been scheduled or until they cannot be scheduled anymore in the same position they were when the rescheduling process began. In the latter case, the new request is removed, the requests are scheduled unmodified and the rescheduling fails. The bandwidth for every request is modified by the following rule: the system decreases the current bandwidth to the value that is the minimum between the needed amount of bandwidth for the new request and half of the current bandwidth used by the request.

### 6.3. Rescheduling for movable requests

The requests that are suitable for this step of the rescheduling process must follow these rules:
- they have to be in the SCHEDULED status
- they must have a priority lower than the priority of the new request
- they must have one of the following time constraints: NONE or NOTBEFORE. Only these types of time constraints are used because they imply that the request was scheduled using a last fit policy and they can be moved to a later start transfer moment without breaking the user's demands.

The system computes a list consisting of requests that follow the above conditions, sorted by the next parameters: the priority (ascending sort), the number of times the request was rescheduled (ascending sort), the size of the allocated bandwidth (descending sort). This sorting method is used to move in the first place the requests that have low priority, that were not modified a lot of times and that have a large amount of allocated bandwidth (these requests are more likely to free enough bandwidth for the new request and to allow the other requests to be scheduled unmodified). The requests belonging to the list are removed from the scheduler (but not from the system's scheduled requests list) one by one. After every removal the system tries to schedule the new request. The operation ends when all the requests have been removed or the new request is scheduled successfully. If the new request could not be scheduled, the removed requests are not reinserted into the scheduler until the new request goes through all the steps of the rescheduling process.

Finally, the removed requests are added back to the scheduler. If the new start time allocated is different from the previous one, the request will be deleted from the system and the corresponding waiting queue for scheduled requests and a new request will be added with the newly computed parameters.

### 6.4. Dynamic rescheduling

The scheduler needs to dynamically modify its requests in one of the following two situations:
- when a transfer finishes faster than its predicted duration
- when a transfer finishes later than its predicted duration.

In the first case, the transfer is automatically removed from the scheduler when it ends, in order to free the bandwidth allocated for its entire duration.

Every transfer is inserted at its start moment in a waiting queue. When its allocated duration is due, the transfer is extracted from the queue and its status is inspected. If it is still in the RUNNING status, this means that it will run more time that it was supposed to. In this case, the scheduler automatically adds a bandwidth reservation with the highest possible priority at the moment when the transfer should have been finished, which allows the transfer to extend its execution over its due time.

This approach has two advantages:
- enables the rescheduling of the requests that were placed after the running transfer if their total bandwidth exceeds the available one
- prevents the scheduling of new requests that could overlap the extended transfer.

## 7. Tests

### 7.1. Testing environment
The MonALISA[2] service was installed on a set of nodes, presented in Figure 2:

| Nr. | Name | Links | Available |
|---|---|---|---|
| 1 | gs | 1 | yes |
| 2 | tschedUPB1 | 1 | yes |
| 3 | tschedUPB2 | 2 | yes |

**Figure 2. System nodes**

The nodes are connected using the links defined below (Figure 3):

| Nr. | Name | Master | From | To | Available | Bandwidth |
|---|---|---|---|---|---|---|
| 1 | link2 | gs | gs | tschedUPB2 | yes | 100 mbps |
| 2 | link1 | tschedUPB1 | tschedUPB1 | tschedUPB2 | yes | 50 mbps |

**Figure 3. System links**

The network topology was defined locally for every farm. The links are considered to be dedicated, having no other traffic than the one initiated by the scheduler.

To facilitate the visualization of the requests and of the way they are kept into the scheduler, the system uses ApMon[14] to send scheduler data to the local MonALISA farm. Every time a new request is inserted into the system, the monitoring module sends the current state of the scheduler to the farm.

The tests consisted in several types of transfers and the result returned by the scheduler. The monitoring information is from the GUI MonALISA client, and shows the data reported by the scheduler to the local MonALISA service.

### 7.2. Transfer that triggers bandwidth modification rescheduling

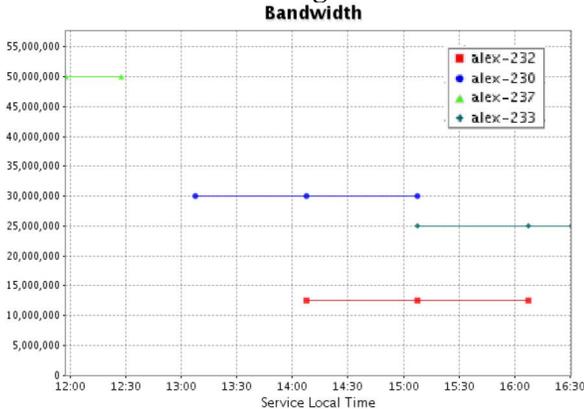

**Figure 4. Scheduler initial state**

The chart in Figure 4 shows four transfers recorded in the scheduler, their scheduled positions in time and their allocated bandwidth.

We try to add a bandwidth reservation that overlaps the first scheduled transfer (alex-237) and has a higher priority. The first transfer (alex-237) has been inserted without a bandwidth specified by the user, and with a priority equal to 1. The scheduler allocated it the maximum available bandwidth, 50Mbps. The new request (alex-238) cannot be scheduled with its specified bandwidth. As a consequence, the bandwidth of the first scheduled request (that has a lower priority) is modified to 12.5Mbps and the new request can now fit into its intended place.

Figure 5 shows the new request (alex-238) inserted with its 30Mbps bandwidth and the transfer with the modified bandwidth of 12.5 Mbps (alex-237).

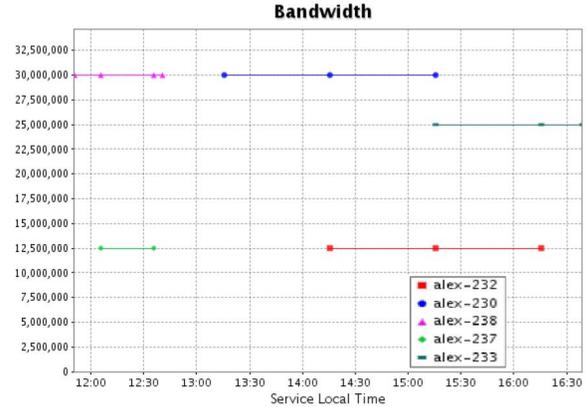

**Figure 5. Scheduler state after inserting alex-238**

### 7.3. Dynamic bandwidth modification for running requests

For a path that has the maximum bandwidth 50 Mbps, we add a request that has no bandwidth specified by the user, alex-145. The path is free, so the request is scheduled with the maximum possible bandwidth, 50Mbps. We try to add a second request, alex-149, again with no bandwidth specified by the user, but with a higher priority (2) and a time constraint: the request should not begin after 12:27 12.06.2008.

The request cannot be scheduled with a bandwidth equal to the maximum value, 50Mbps, and the system tries again to schedule it with a bandwidth equal to half the maximum bandwidth, 25Mbps. Because of the running request, the scheduling process fails again. As a consequence, the bandwidth of the running request (alex-145) is dynamically modified to 25Mbps and the new request (alex-149) gets an assigned bandwidth of 25Mbps.

We add a third request, alex-150, with a bandwidth of 20Mbps, but with a higher priority (5) than the previous two requests and a time constraint: the transfer should not begin after 12:30 12.06.2008. The transfer cannot be scheduled, and given its high priority, the other two running transfers go through another bandwidth modification process: their bandwidth drops from 25Mbps to 12.4 Mbps. The chart in Figure 6 shows the bandwidth evolution of the three requests and the sum of their bandwidths, which has to be lower than 50Mbps.

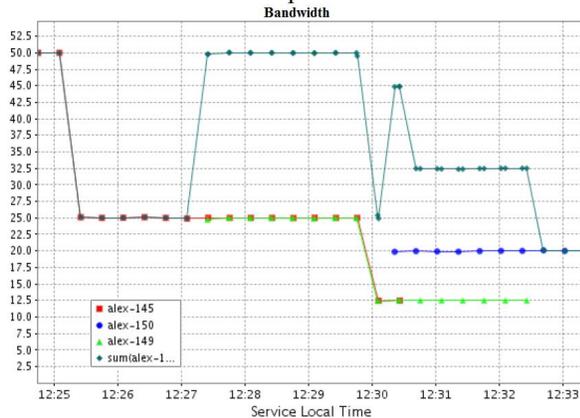

**Figure 6. Bandwidth values for running requests**

## 8. Conclusions

The proposed file transfer scheduler takes care of the different aspects in transferring data in a dynamic environment. It offers a reservation system for user transfer requests that allocates them transfer times and bandwidth, according to the constraints the user specified for the requests.

The user has the possibility to reserve a certain time moment for his request, or simply make a bandwidth reservation in order to be sure that the desired bandwidth is available. If the transfer is successfully scheduled, the user that submitted the request knows when the transfer is going to take place and its duration. The system will always keep the parameters stated by the user at their specified value. When the bandwidth or the desired start time are not important for the user, they are allocated by the scheduler, in the best possible way to make use of the available bandwidth and to allow other requests to run concurrently. The scheduler also offers the possibility of establishing a hierarchy between requests, by assigning them different priorities. If a high priority request cannot be scheduled, the system tries to dynamically adjust the parameters for low priority requests in order to accommodate the new one.

The new features that our algorithm brings compared to other approaches can be summarized as follows: it introduces for the user the possibility to let the system decide the value of the bandwidth that the transfer will use; it establishes a hierarchy for the transfers, enabling the rescheduling of the low priority requests, in terms of dynamically reducing the allocated bandwidth for running and scheduled requests, and moving the scheduled requests that have no time constraints; it adds to the running requests that exceed their allocated time period the means to automatically trigger the rescheduling of subsequent transfers and continue their execution without affecting the scheduling process.

The possible improvements to the scheduling algorithm can be related to the next ideas:
- the scheduler can integrate MonALISA monitoring information to provide accurate measurements of the available bandwidth
- for the duration of the file transfers, the scheduler could use a duration prediction module
- the system could tackle the problem of transfers with dependencies.

## 9. References

[1] C. Cirstoiu, "Optimization Framework for Data Intensive Applications in Large Scale Distributed Systems", PhD. dissertation, University "Politehnica" of Bucharest, 2008.
[2] "MonALISA" http://monalisa.caltech.edu/
[3] H.B. Newman, I.C.Legrand, P. Galvez, R. Voicu, C. Cirstoiu, "MonALISA - A Distributed Monitoring Service Architecture", *CHEP03*, La Jolla, California, 2003
[4] Y. Zhu, "A Survey on Grid Scheduling Systems", Technical Report, Computer Science Department of Hong Kong, University of Science and Technology, 2003
http://oslab.khu.ac.kr/mgrid/resources/rao_ZHU_Yanmin_Survey_Report.pdf
[5] Reliable File Transfer (RFT),
 http://www.globus.org/toolkit/data/rft/
[6] A. Popescu, "Integrated transfer scheduling system with reservations in Grid environments", Diploma Project, University "Politehnica" of Bucharest, 2007.
[7] „File Transfer Service", http://egee-jra1-dm.web.cern.ch/egee-jra1-dm/FTS/default.htm
[8] „EGEE gLite User's Guide GLITE FILE TRANSFER SERVICE – CLI", 2005, https://edms.cern.ch/file/591792/1/EGEE-TECH-591792-Transfer-CLI-v1.0.pdf
[9] T. Kosar and M. Livny, "Stork: Making data placement a first class citizen in the Grid", *Proceedings of the 24th International Conference on Distributed Computing Systems,* 2004, http://www.cs.wisc.edu/condor/stork/papers/stork-icdcs2004.pdf
[10] G.A. Stewart, G. McCance, "Grid Data Management: Reliable File Transfer Services' Performance", *CHEP06*, Mumbai, India, February 2006
[11] Ian Foster, Carl Kesselman, "Globus: A Metacomputing Infrastructure Toolkit", *International Journal of Supercomputing Applications*, 1997
[12] "The Globus toolkit webpage", www.globus.org/toolkit/
[13] "Globus GridFTP",
http://www.globus.org/toolkit/docs/4.0/data/gridftp/index.pdf
[14] ApMon User Guide,http://monalisa.cacr.caltech.edu/monalisa__Documentation__ApMon_User_Guide.htm